\documentstyle[a4,epsfig,12pt,rotating,clic]{article}
\begin{document}

\date{31 January, 2001}

\title{Physics Signatures at CLIC}

\author{Marco Battaglia}

\def\scdinstitute{CERN - PS Division}

\def\scdnote{LC-PHSM-2001-072-CLIC}

\def\scdconference{Based on material presented at the\\ 
1$^{st}$ Open Meeting for CLIC Physics Studies \\ CERN, 25 - 26 May, 2000}

\long\def\scdabstract{
A set of signatures for physics processes of potential interests for the 
CLIC programme at $\sqrt{s} =$ 1 - 5~TeV are discussed. These
signatures, that may correspond to the manifestation of different scenarios
of new physics as well to Standard Model precision tests, are 
proposed as benchmarks for the optimisation of the CLIC accelerator parameters
and for a first definition of the required detector response. 
}

\titlepage

\pagebreak

\setcounter{page}{1}
   
\section{Introduction}

The {\sc Clic} concept aims for a linear collider providing collisions at 
centre-of-mass energies in the range 1~TeV $< \sqrt{s} <$ 5~TeV, 
with a luminosity of $10^{35}$~cm$^{-2}$~s$^{-1}$ at $\sqrt{s}$ = 3~TeV, 
based on the two-beam acceleration scheme~\cite{clic}. While opening a new 
domain for experimentation at $e^+e^-$ colliders, {\sc Clic}
also presents new challenges compared to any other linear collider project,
due to its different regime of operation, thus requiring a significant effort
in the optimisation of its parameters and assessing the implications of the
machine induced backgrounds on its physics potential.

The definition of its physics programme still requires essential data that is
likely to become available only after the first years of {\sc Lhc} operation 
and, possibly, also the results from a lower energy linear collider such as 
{\sc Tesla}~\cite{tesla_cdr}. At present we have to envisage several possible 
scenarios for the fundamental questions to be addressed by HEP experiments in 
the second decade of the new century. 

In this note a set of benchmark physics signatures for the optimisation of the
{\sc Clic} parameters are proposed. Each of these signatures may signal 
the manifestation of different scenarios of new physics, possibly beyond those 
that we are able to define today. The optimisation of the machine design and 
the study of its potential for each of them will also guarantee a more general 
applicability to other processes, characterised by similar topologies.

Furthermore, while considering experimentation at a multi-TeV collider, it is 
interesting to verify to which extent extrapolations of the experimental 
techniques, successfully developed at {\sc Lep} and being refined in the
studies for the {\sc Tesla} project, are still applicable. This has important
consequences on the requirements for the experimental conditions at the 
{\sc Clic} interaction region and for the definition of the {\sc Clic} physics
potential.

\section{Physics Signatures}

These considerations have motivated a preliminary investigation of 
physics signatures at {\sc Clic} that are characteristic of one or more 
physics scenarios of interest and relevant to specific aspects of the 
accelerator performances and/or of the detector response. 
Four physics signatures have been identified:

\begin{enumerate}
\item{Resonance Scan}
\item{Electro-weak Fits}
\item{Multi-Jet Final States}
\item{Missing Energy and Forward Processes}
\end{enumerate}

\begin{figure}[h!]
\begin{center}
\vspace*{-0.75cm}
\epsfig{file=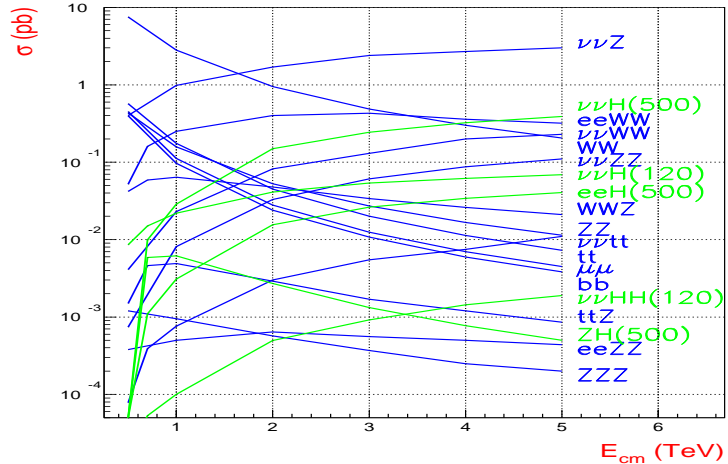,height=7.0cm,width=11.0cm}
\vspace*{-0.75cm}
\end{center}
\caption{\sl Cross sections for SM processes at $\sqrt{s}$ = 1-5~TeV.}
\label{fig:xs}
\end{figure}

In the following, these are discussed in some details on the basis of the 
first results of dedicated simulations of several relevant physics processes.
The cross sections have been estimated using the {\sc CompHep} 
program~\cite{comphep} and the simulated events generated using the 
{\sc Pythia-6} Monte Carlo~\cite{pythia} (see Figure~\ref{fig:xs}).
Events have been passed through either a parametric smearing using the 
{\sc Simdet} program~\cite{simdet} or a full {\sc Geant-3}~\cite{geant3} 
based simulation. In both cases
the detector design and its performances have been assumed to correspond to 
those of the detector designed for the {\sc Tesla} linear collider Conceptual 
Design Report~\cite{tesla_cdr}. In view of the different energy and 
backgrounds here considered for {\sc Clic}, the strength of the magnetic 
field and the radius of the detector layer closer to the beam pipe have been 
changed from 3~T to 4-6~T and from 1.5~cm to $\simeq$~3.0~cm respectively. 
\begin{figure}[h!]
\begin{center}
\vspace*{-0.75cm}
\epsfig{file=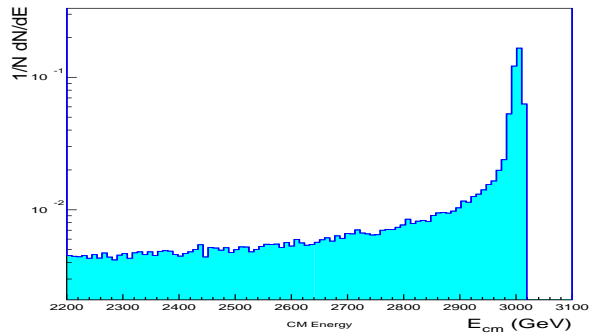,height=5.0cm,width=9.0cm}
\vspace*{-0.75cm}
\end{center}
\caption{\sl The upper portion of the {\sc Clic} luminosity spectrum at 
$\sqrt{s} \simeq$ 3~TeV.}
\end{figure}
The expected luminosity spectrum and $\gamma \gamma$ background have been 
obtained 
from the {\sc GuineaPig} simulation, with the {\sc Clic} reference parameters 
at $\sqrt{s}$ = 3~TeV~\cite{backg}, and interfaced to {\sc Pythia} with the 
{\sc Calypso} and {\sc Hades} programs respectively~\cite{hades}.

\subsection{Resonance Scan}

The most striking manifestation of new physics in the multi-TeV region will 
come from the sudden increase of the $e^+e^- \rightarrow f \bar f$ cross 
section signalling the s-channel production of a new particle. 
There are several
scenarios predicting new resonances in the mass range of interests for 
{\sc Clic}. The first one is the existence of extra gauge bosons such as
a $Z'$ boson. This is common to both GUT-inspired $E_6$ models and to
Left-Right symmetric models. A $Z'$ boson within the kinematical reach of
{\sc Clic} should be already observable at the {\sc Lhc} and {\sc Clic} will
concentrate on the accurate measurements of its properties by a direct
resonance scan and by the
electro-weak fits discussed in the next section. It is interesting to note 
that a recent re-analysis of data on atomic parity violation shows that the 
observed 2.8~$\sigma$ deviation from the SM prediction~\cite{apv} can be 
explained by a $Z'$ boson with mass in the range 
0.6~TeV/$c^2$ $< M(Z') <$ 1.5~TeV/$c^2$ that is compatible with the present
limits~\cite{decurtis}.  

New resonances are also predicted by new theories of gravity with extra 
dimensions in the form of Kaluza-Klein graviton and gauge boson excitations. 
Their phenomenology at the TeV scale has been analysed~\cite{rizzo1,rizzo2} 
and the results can be extended to the case of the {\sc Clic} multi-TeV 
collider.
Models of dynamical electro-weak symmetry breaking also predict the existence 
of new resonances in the TeV region. In particular the degenerate BESS model 
introduces a pair of narrow and nearly degenerate vector and an axial-vector 
resonances~\cite{dbess}. The study of such resonances will have to establish 
their nature, to accurately determine their mass and width and to disentangle
a wide resonance from the superposition of two nearly degenerate states.

\begin{figure}[h!]
\begin{center}
\vspace*{-0.75cm}
\epsfig{file=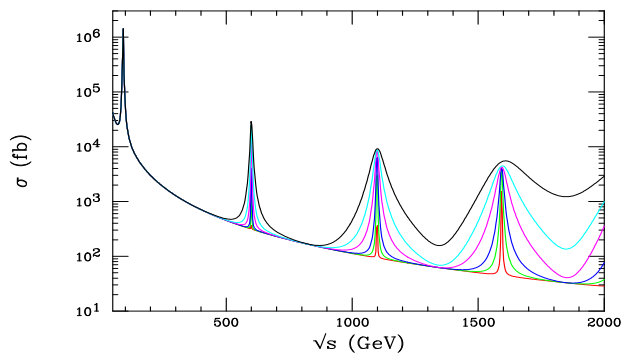,width=5.5cm,height=9.0cm,angle=90}
\vspace*{-0.75cm}
\end{center}
\caption{\sl $\sigma(e^+e^- \rightarrow \mu^+ \mu^-)$ vs. 
$\sqrt{s}$ for a fixed spectrum of KK resonances with different width
$\Gamma_{KK} =  f(c) M^3_{KK}$ in the RS model~\cite{randall} 
(from~\cite{rizzo1}).}
\label{fig:kk}
\end{figure}

Most relevant to the resonance scan signature are the characteristics of the 
luminosity spectrum and the ability to perform jet flavour identification. 

\begin{figure}[h!]
\begin{center}
\vspace*{-0.75cm}
\begin{tabular}{c c}
\epsfig{file=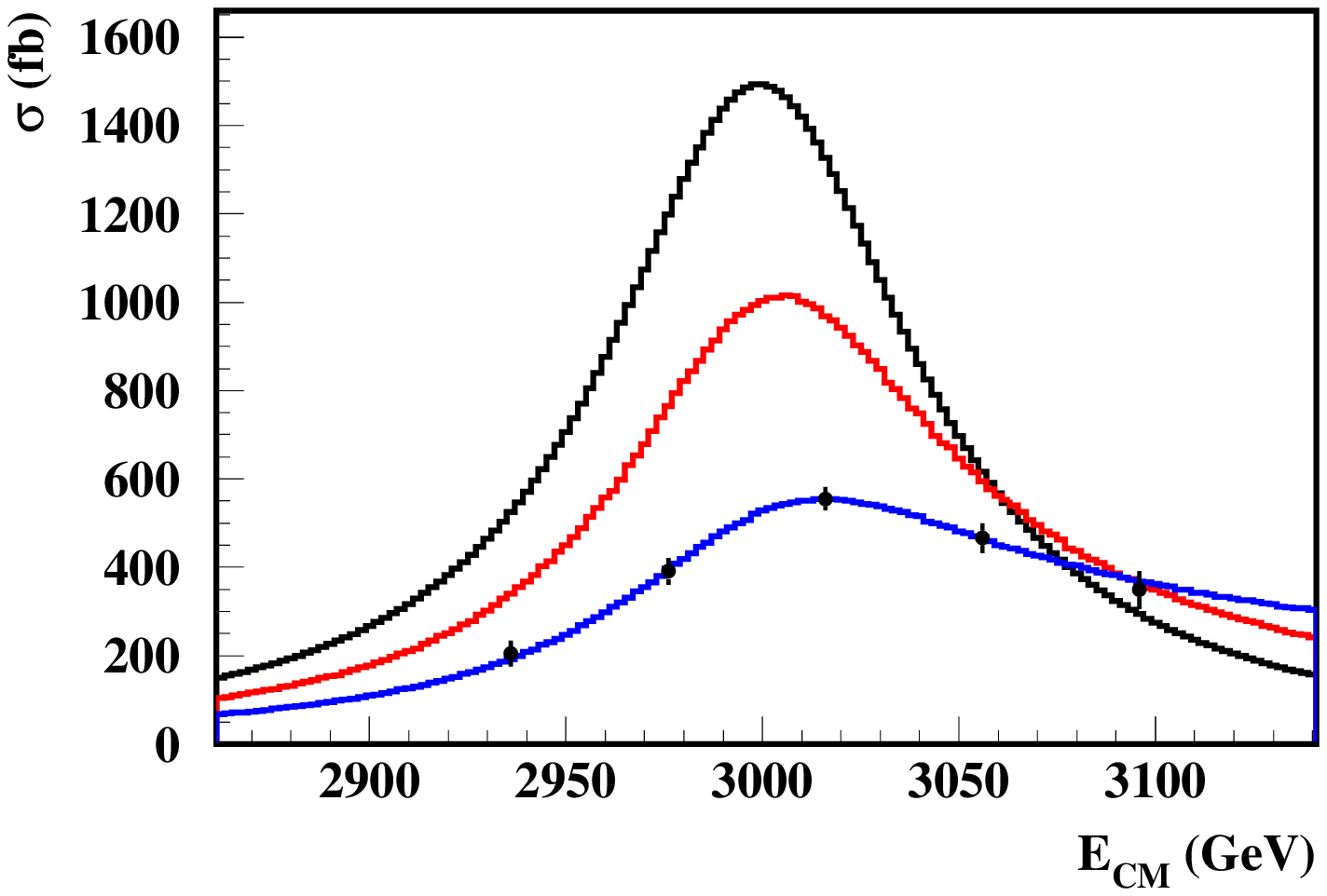,width=6.5cm,height=6.0cm,clip} &
\epsfig{file=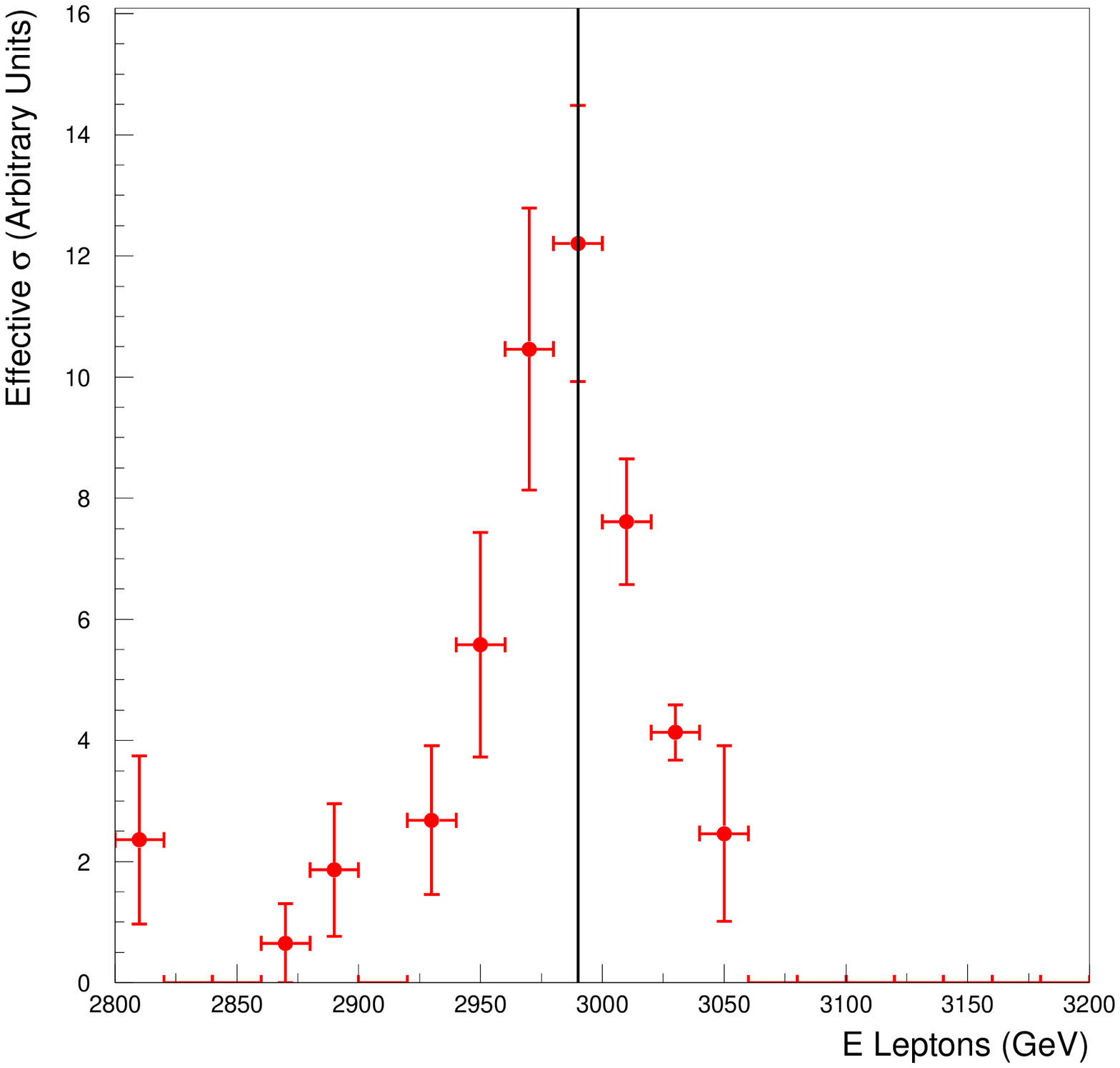,width=6.5cm,
height=6.0cm,clip} \\
\end{tabular}
\vspace*{-0.75cm}
\end{center}
\caption{\sl The $Z'$ resonance scan obtained with energy scan (left) and
auto-scan (right). In the left plot, the Born production cross-section, the 
cross section with ISR included and that accounting for the 
luminosity spectrum and the tagging criteria are shown.} 
\label{fig:zscan}
\end{figure}

A preliminary study of on-shell $Z'$ production at CLIC has been performed 
already based on parametric smearing. 
$e^+e^- \rightarrow Z'$ events have been generated for 
$M(Z')$ = 3.0~TeV/$c^2$ including ISR, beam energy spectrum 
and $\gamma \gamma$ backgrounds assuming SM-like couplings, corresponding to
a width $\Gamma_{Z'} \simeq$ 90~GeV/$c^2$. Due to the 
large signal cross-section over a flat background and the large {\sc Clic}
luminosity, a precise study of the mass, width and couplings can be performed.
The mass and width of resonance can be determined by performing either an 
energy scan, like the $Z^0$ scan performed at {\sc Lep}/{\sc Slc} and foreseen
for the $t \bar t$ threshold, or an auto-scan, by tuning the collision energy 
just beyond the top of the resonance and profiting  of the long tail of the 
luminosity spectrum to probe the resonance peak. For the first method
both di-jet and di-lepton final states can be considered, while for the 
auto-scan only $\mu^+ \mu^-$ final states may provide with the necessary 
accuracy for the $Z'$ energy.  Preliminary results are shown in 
Figure~\ref{fig:zscan}. Assuming an integrated luminosity of 1000~fb$^{-1}$,
corresponding to one year of {\sc Clic} running at its nominal luminosity,
a relative statistical accuracy of $\pm~5~\times~10^{-5}$ for the mass and of
$\pm~3~\times~10^{-3}$ for the total width can be obtained.  

\subsection{Electro-weak Fits}

If a new resonance, such as a $Z'$, will have been found at the {\sc Lhc} 
or at a lower energy linear collider, the {\sc Clic} physics programme will 
be largely focused on the precise study of its nature via the measurement of 
its fundamental properties.
This will closely follow the {\sc Lep} physics programme at the $Z^0$ peak.
The main issue in these studies will be the determination of the nature of the
observed resonance. A precise study of its couplings to up- and down-type 
quarks and of forward-backward asymmetries may be instrumental in this 
program. As an example, the nearly degenerate Kaluza-Klein excitations of the 
$Z$ and $\gamma$~\cite{rizzo3} or the $L_3$ and $R_3$ resonances of the BESS 
model need to be distinguished from a $Z'$ boson. 

Even if such a resonance is not found, precise electro-weak data for
$\sigma (e^+e^- \rightarrow f \bar f)$, 
$A^{f \bar f}_{FB}$ and $A^{f \bar f}_{LR}$ (with $f \bar f$ = $\ell^+ \ell^-$,
$c \bar c$, $b \bar b$, $t \bar t$) at $\sqrt{s}$ = 1 - 5~TeV may give 
indirect evidence of new phenomena such as an heavier $Z'$ well beyond 
the {\sc Clic} kinematical limit, extra-dimensions or contact interactions. 
It is therefore important to ensure that accurate electro-weak data can be 
obtained.

The main experimental requirements highlighted by electro-weak fits are the 
identification of the $e^+e^- \rightarrow f \bar f$ from $\gamma \gamma 
\rightarrow {\mathrm{hadrons}}$ final states, the ability to reconstruct 
and tag the fermions down to small polar angles, in presence of a significant 
$\gamma \gamma \rightarrow$~hadrons background, the reconstruction of the 
colliding $e^+e^-$ rest frame in presence of large beamstrahlung and the 
possibility to obtain highly polarised $e^-$ and $e^+$ beams.

Good examples of the {\sc Clic} potential are given by the sensitivity
to a heavy $Z'$ and to extra-dimensions. Direct or indirect
evidence of an additional neutral gauge boson will require a detailed study of
its couplings to determine the underlying extended gauge structure.
Extrapolating the results of a previous study on extra dimensions based on the
forward-backward asymmetry $A_{fb}$ in $e^+e^- \rightarrow b \bar 
b$~\cite{joanne}, the {\sc Clic} sensitivity for $M_S$, 
representing the scale at which gravity becomes strong, is given by
$\simeq 7 \times \sqrt{s}$~(TeV) if the $b \bar b$ final state can be
identified efficiently. 

\begin{table}[h!]
\begin{center}

\caption{\sl $b \bar b$ production cross-section from 
$\sqrt{s} = M(Z^0)$ up to 3~TeV.}

\vspace{0.25cm}

\begin{tabular}{|l|c|c|c|c|}
\hline 
$\sqrt{s}$ (TeV)  & 0.09 & 0.5 & 0.8 & {\bf 3.0} \\
\hline \hline 
$\sigma_{b \bar b}$ (pb)& 9160 & 0.4 & 0.15 & {\bf 0.012} \\
$\int L$            & & 500 pb$^{-1}$ & 500 pb$^{-1}$ & 
{\bf 5000 fb$^{-1}$} \\
$N_{b \bar b}$ & 900k & 200k & 75k & {\bf 60k} \\
\hline     
\end{tabular}
\end{center}
\label{table:bbxsec}
\end{table}

Therefore, the identification of $b \bar b$ and $c \bar c$ final states 
represents an important benchmark process also at {\sc Clic}. 
At 3~TeV $\sigma(e^+e^- \rightarrow b \bar b)$ is only of the order of 
0.01~pb but the heavy hadrons receive a significant boost to acquire decay 
distances in space of several centimetres, i.e. an order of magnitude larger 
compared to those at {\sc Lep} energies.
\begin{table}[h!]
\begin{center}

\caption{\sl Average decay distance in space for $B$ hadrons 
at different $\sqrt{s}$} 

\vspace{0.25cm}

\begin{tabular}{|l|c|c|c|c|c|}
\hline
$\sqrt{s}$ (TeV) & 0.09  & 0.2 & 0.35 & 0.5 & {\bf 3.0} \\ \hline
Process          & $Z^0$ & $HZ$ & $HZ$ & $HZ$ & {$H^+H^-$} $|$ 
{$b \bar b$} \\
\hline \hline
$d_{space}$ (cm) & 0.3 & 0.3 & 0.7 & 0.85 & ~~~{\bf 2.5}~~~~$|$ 
{\bf 9.0}\\
\hline
\end{tabular}
\end{center}
\label{table:bbdec}
\end{table}
This has important experimental consequences. The long decay distance of
$B$ hadrons, exceeding the anticipated radii of the first layers of the 
Vertex Tracker, suggests a tagging algorithm based on the topological tag of 
secondary vertices rather than on large track impact parameters as performed
at {\sc Lep}. A beauty tag based on the change in charged multiplicity 
detected through the layers of a Si Vertex Tracker, that applies
to {\sc Clic} a concept originally developed in charm photo-production 
experiments~\cite{na1}, has been proposed~\cite{vertex2000}.
Such a tag can be part of the track pattern recognition that still needs to 
be studied in full details in the challenging environment produced by the 
high jet collimation and background hits from pair and 
$\gamma \gamma \rightarrow {\mathrm{hadrons}}$ production. This represents an 
important benchmark for the design of the vertex and main trackers. 
Different silicon pixel technologies have been proposed for the Vertex Tracker
at {\sc Tesla} and {\sc Nlc}~\cite{vtx}. Assuming a 25~ns time stamping 
capability, such as for
the LHC hybrid pixel sensors, the Vertex Tracker integrates about 30~bunch
crossing corresponding to an hit density due to pairs of 
$\simeq$~0.17~mm$^{-1}$ at 3.0~cm that becomes $\simeq$~0.8~mm$^{-1}$ if the 
full train of 154 bunches is integrated~\cite{backg}. These rates appear to be 
manageable for a Vertex Tracker with redundant radial measurements and small 
enough cell size.

\begin{figure}[ht!]
\begin{center}
\vspace*{-0.75cm}
\begin{tabular}{c c}
\epsfig{file=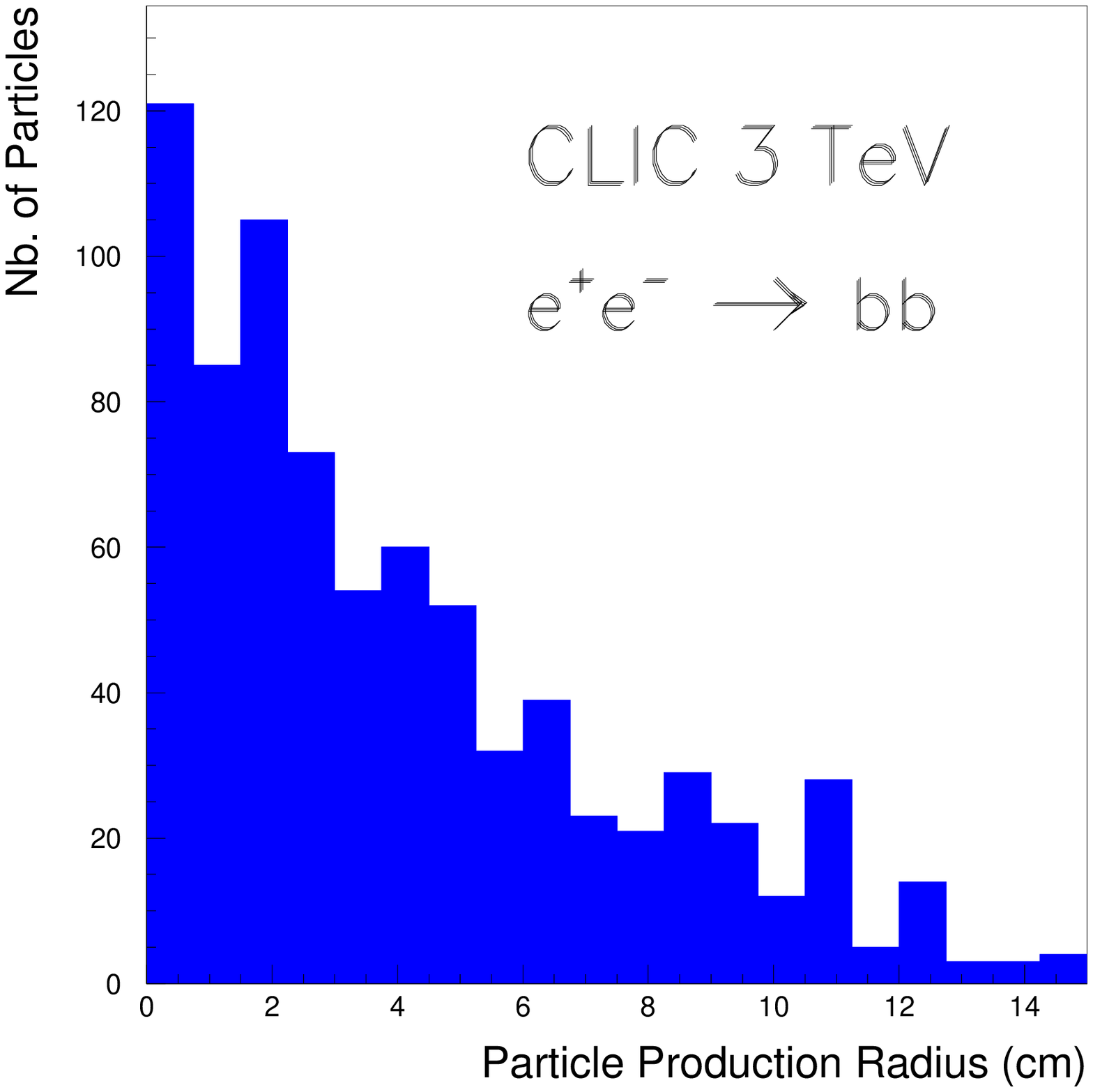,height=5.5cm,width=6.5cm} &
\epsfig{file=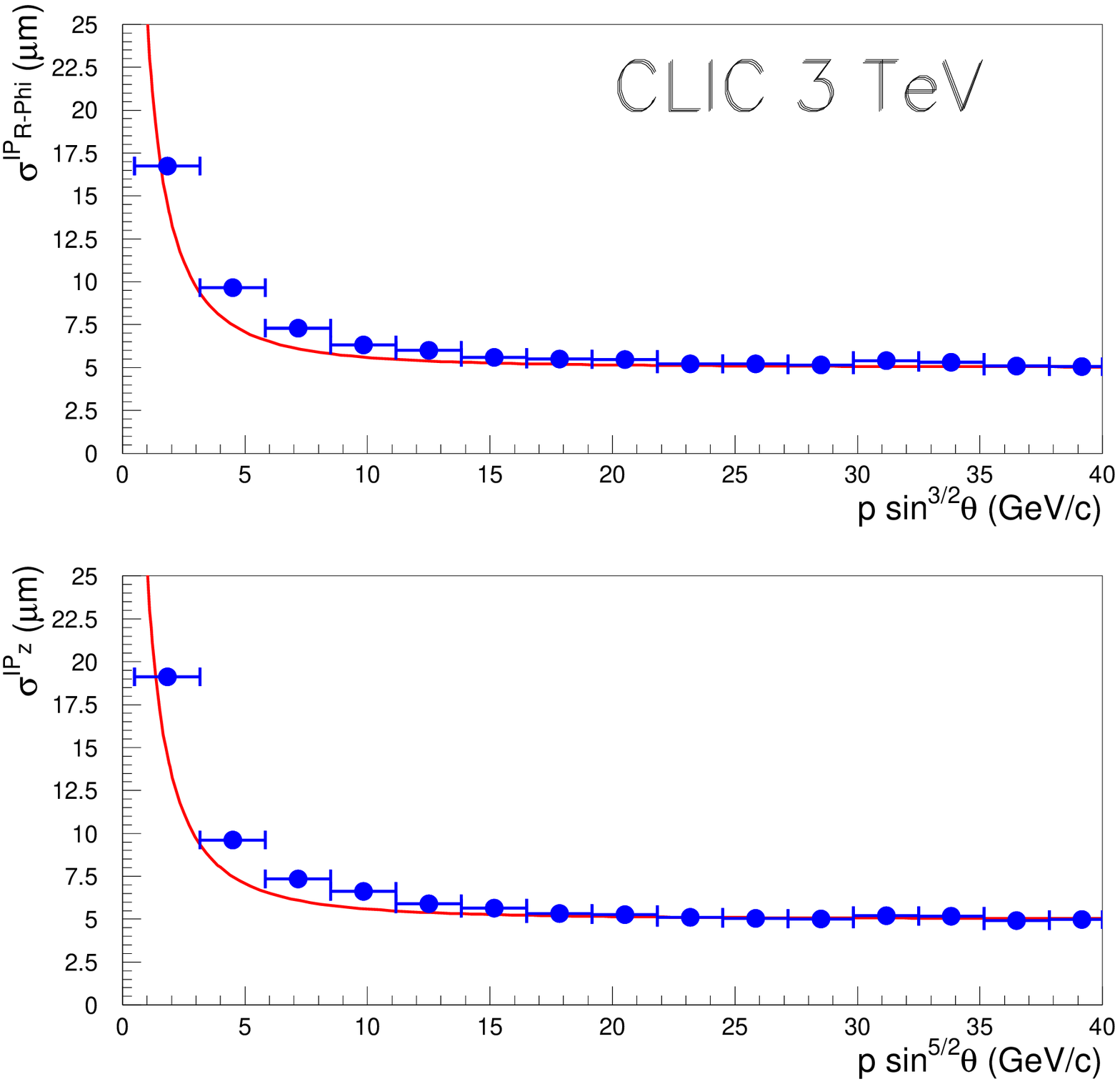,height=5.5cm,width=6.5cm} \\
\epsfig{file=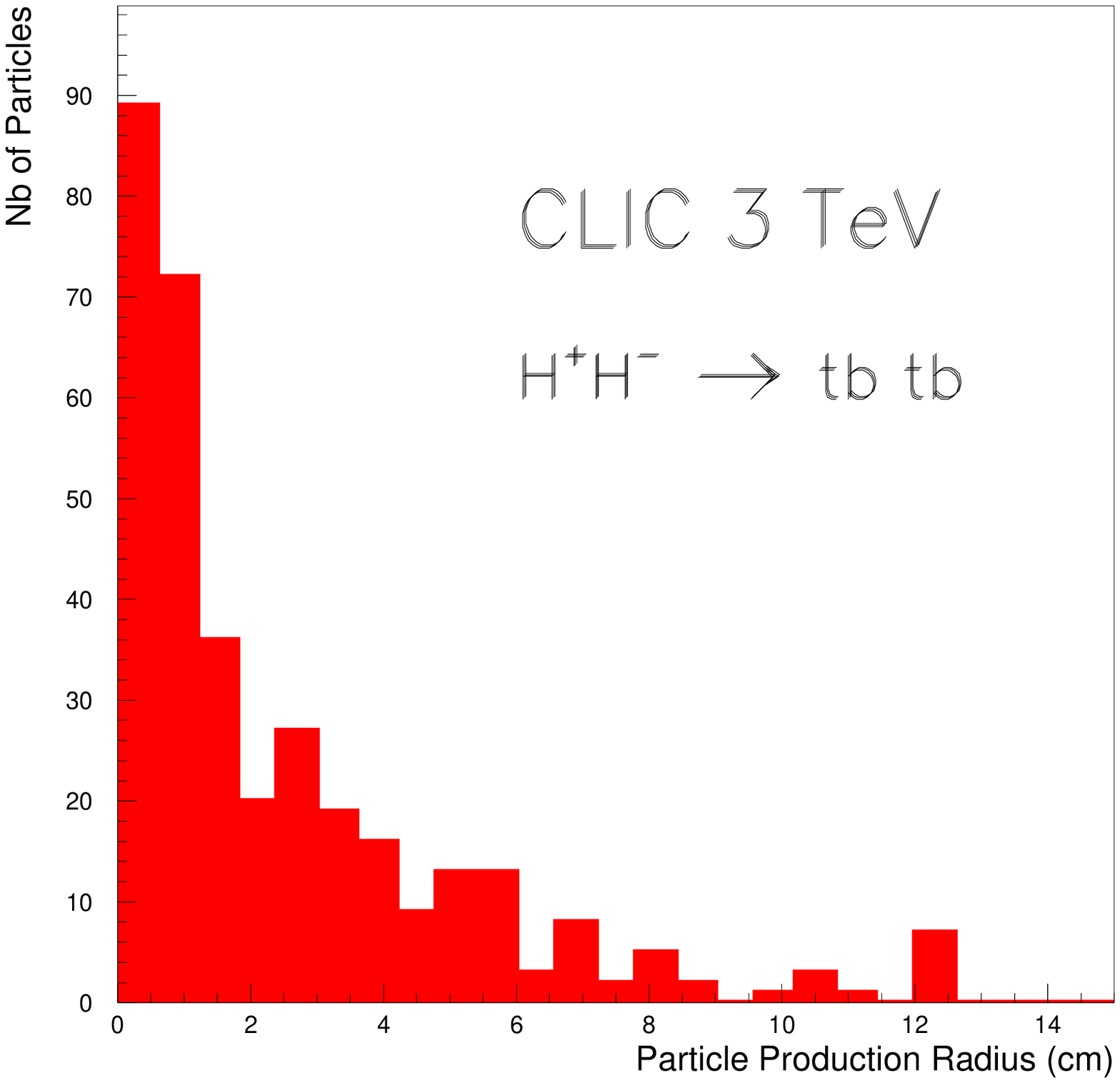,height=5.5cm,width=6.5cm} &
\epsfig{file=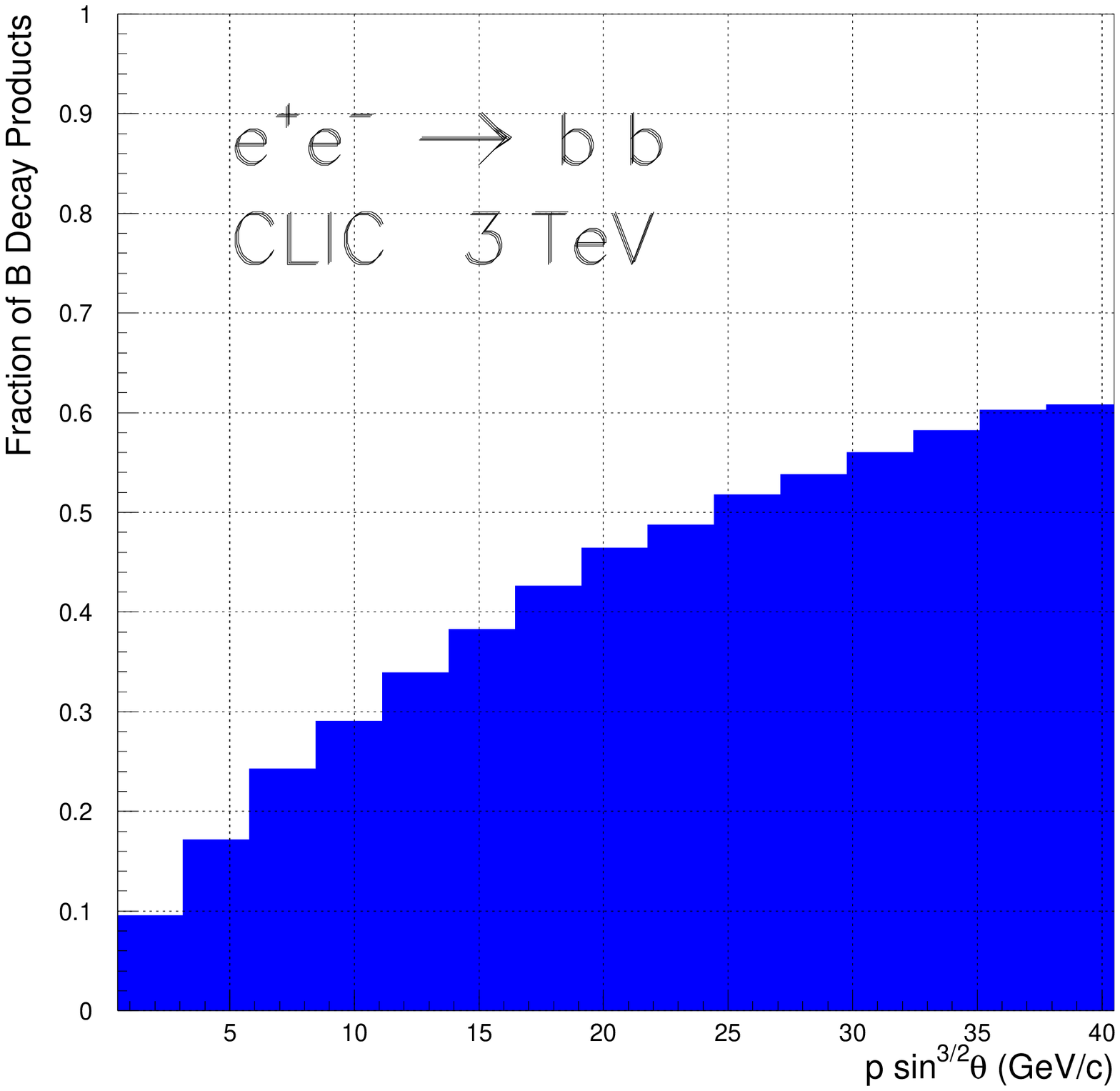,height=5.5cm,width=6.5cm} \\
\end{tabular}
\vspace*{-0.75cm}
\end{center}
\caption{\sl Distributions of the decay distances for $B$ hadrons
produced in $e^+e^- \rightarrow b \bar b$ (upper left) and $e^+e^-
\rightarrow H^+H^- \rightarrow t \bar b \bar t b$ decays (lower left),
impact parameter resolution expected for a Vertex Tracker at
{\sc Clic} plotted as a function of the particle momentum (upper right) and
fraction of secondary particles with momentum below a given value 
(upper left) in $b \bar b$ events.}
\label{fig:ip}
\end{figure}
The resolution for extrapolating a particle track to its production point
has been estimated assuming the first sensitive layer at 3.0~cm radius,
a single point resolution of 8~$\mu$m and 250~$\mu$m sensor thickness that
represent realistic assumptions for a Vertex Tracker based on hybrid pixel
sensors~\cite{vtx}. 
The result is shown in ~Figure~\ref{fig:ip} as a function of the particle 
momentum.

\subsection{Multi-Jet Final States}

Multi-jet final states are a common signature for many physics processes
from  heavy Higgs pair production to $WW$ scattering.
Evolving from the $Z^0$ peak to multi-TeV c.m.\ collisions, the number of 
jets increases due to gluon radiation and the crossing of thresholds
for multi-fermion final state production. 

\begin{table}[h!]
\begin{center}

\caption{\sl Average reconstructed jet multiplicity in hadronic events at 
different $\sqrt{s}$ energies.}

\vspace{0.25cm}

\begin{tabular}{|l|r|r|r|r|r|r|r|}
\hline
$\sqrt{s}$ (TeV) & 0.09 & 0.20 & 0.5 & 0.8 & 3.0 & 5.0 \\ 
\hline \hline  
$<N_{Jets}>$     & 2.8  & 4.2  & 4.8 & 5.3 & 6.4 & 6.7 \\
\hline 
\end{tabular}
\end{center}
\label{tab:jmult}
\end{table}

There are several physics processes of possible interest at {\sc Clic}
energies characterised by multi-jet final states. In the Higgs sector, the 
reconstruction of the Higgs potential, requiring the determination of the 
triple and quartic Higgs couplings through the measurement of double and 
triple Higgs production diagrams, and the study of the Higgs Top-Yukawa 
coupling will be required to complete the Higgs profile investigation and as 
proof of the Higgs mechanism of mass generation. The small values of the 
cross sections for these processes will require the large collision energies 
and integrated luminosities that can be achieved at {\sc Clic}~\cite{vhiggs}. 
Even for a Higgs boson as light as 120~GeV/$c^2$, the $e^+e^- \rightarrow 
\nu_e \bar \nu_e H H$ cross section is only 0.05~fb at $\sqrt{s}$ = 0.8~TeV 
and becomes 0.92~fb at $\sqrt{s}$ = 3.0~TeV. The triple Higgs production is 
suppressed by a factor of $\simeq10^3$. {\sc Clic} could produce enough of 
these events for the triple Higgs couplings to be determined to better than 
10\% and to obtain also information on the elusive quartic coupling.
 
\begin{table}[h!]
\begin{center}

\caption{\sl Examples of Higgs processes with large jet multiplicity.}

\vspace{0.25cm}

\begin{tabular}{|l|c|c|r|c|}
\hline
Process & $\sigma$  & $M_H$ & Events~~  & $N_{Jets}$ \\
        & (fb)      & (GeV/$c^2$) & / 5000 fb$^{-1}$ &   \\
\hline \hline
$t \bar t H \rightarrow WW/ZZ$ ($t \bar t$)  & 0.18 & 500 & 850 & 
10 (12) \\
$ZHH \rightarrow WW/ZZ$ ($t \bar t$) & 0.038 & 500 & 190 & 10 (14) \\
$\nu \bar \nu HH \rightarrow WW/ZZ$  & 0.062 & 500 & 310 & 8 (12) \\
$ZHH \rightarrow b \bar b$ & 0.036 & 120 & 180 & 6 \\
$\nu \bar \nu HH \rightarrow b \bar b$ & 0.94 & 120 & 4700 & 4 \\
$H^+H^- \rightarrow t \bar b ~\bar t b$ &  2.0 & 800 & 10000 & 8 \\
\hline
\end{tabular}
\end{center}
\label{tab:jhiggs}
\end{table}

\begin{figure}[h!]
\begin{center}
\vspace*{-0.75cm}
\begin{tabular}{c c}
\epsfig{file=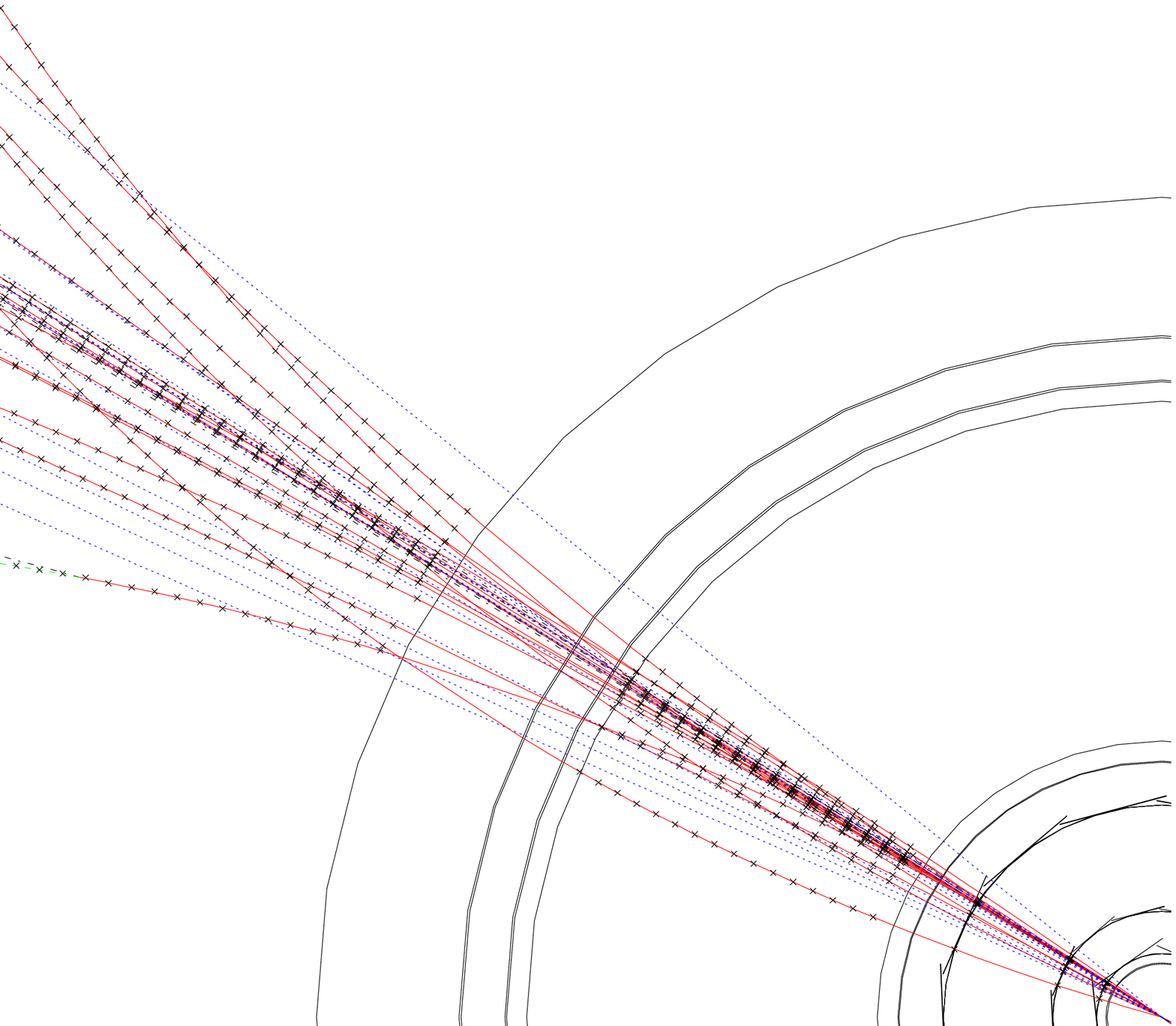,width=5.0cm} &
\epsfig{file=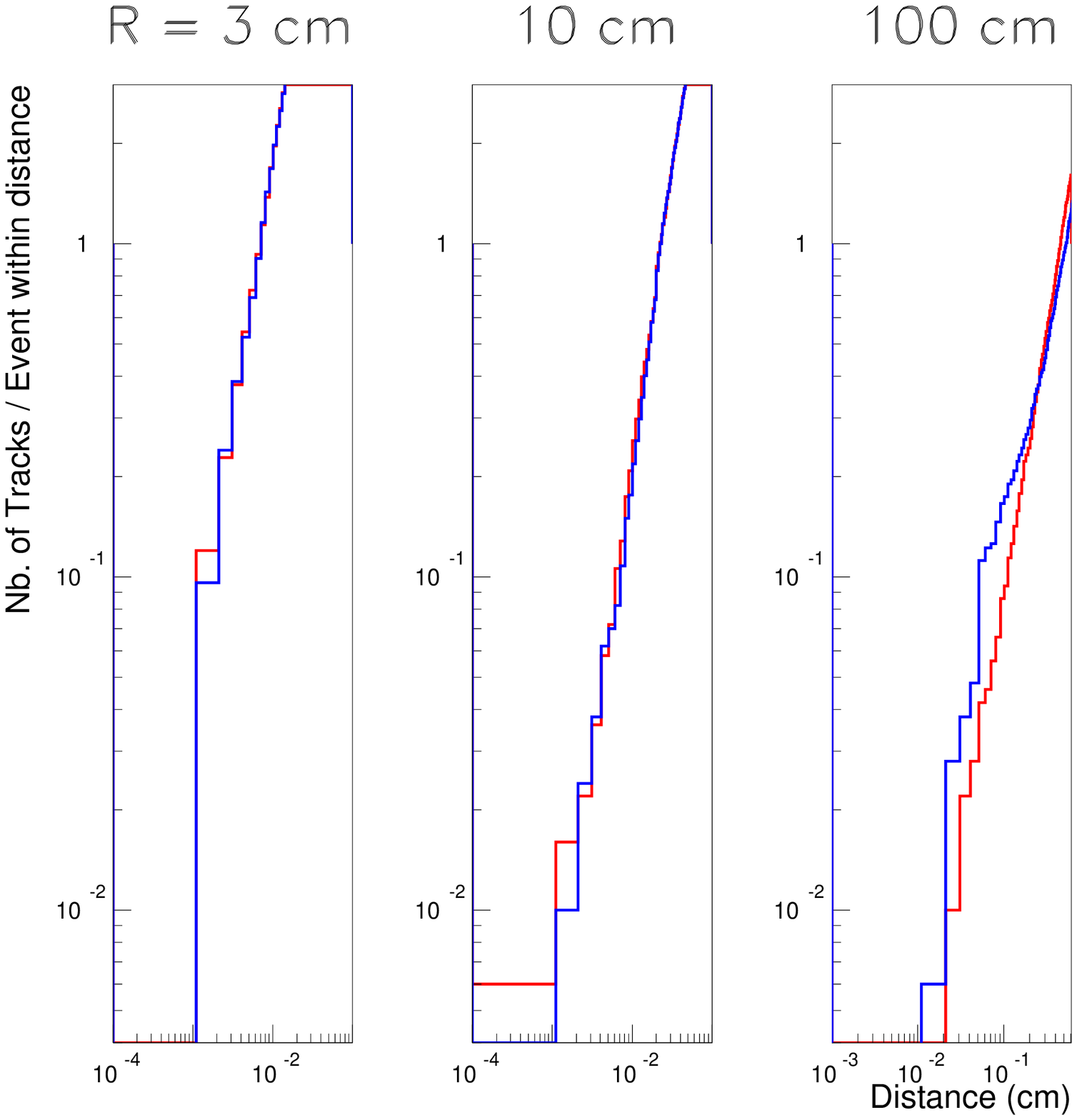,width=8.0cm,height=6.0cm}
\end{tabular}
\vspace*{-0.75cm}
\end{center}
\caption{\sl Magnified view of a di-jet from a $e^+e^- \rightarrow W^+W^-$ 
event at $\sqrt{s}$ = 3~TeV (left) and the distance from the closest track for
charged particle with $p >$ 3~GeV/$c$ at different radial distances (right)
for B = 4~T (dark grey histogram) and 6~T (light grey histogram).}
\label{fig:occup}
\end{figure}

As another example, the heavy Higgs sector in SUSY models may require a 
multi-TeV collider for accessing $e^+e^- \rightarrow H^+H^-$ and $H^0 A^0$ 
pair production.
As indicated in Table~4, these processes are characterised by 
large jet multiplicities that need to be properly handled in order to isolate
the small expected signals.
Multi-jet final states probe the detector performance in terms of two
particle separation and energy flow, while the broad luminosity 
spectrum represents a source of inaccuracy when performing constrained fits
and the large $\gamma \gamma \rightarrow$~hadrons background expected at
{\sc Clic} may confuse the topological event reconstruction adding or smearing
jets in the forward regions.

\begin{figure}[h!]
\begin{center}
\vspace*{-0.75cm}
\begin{tabular}{c c}
\epsfig{file=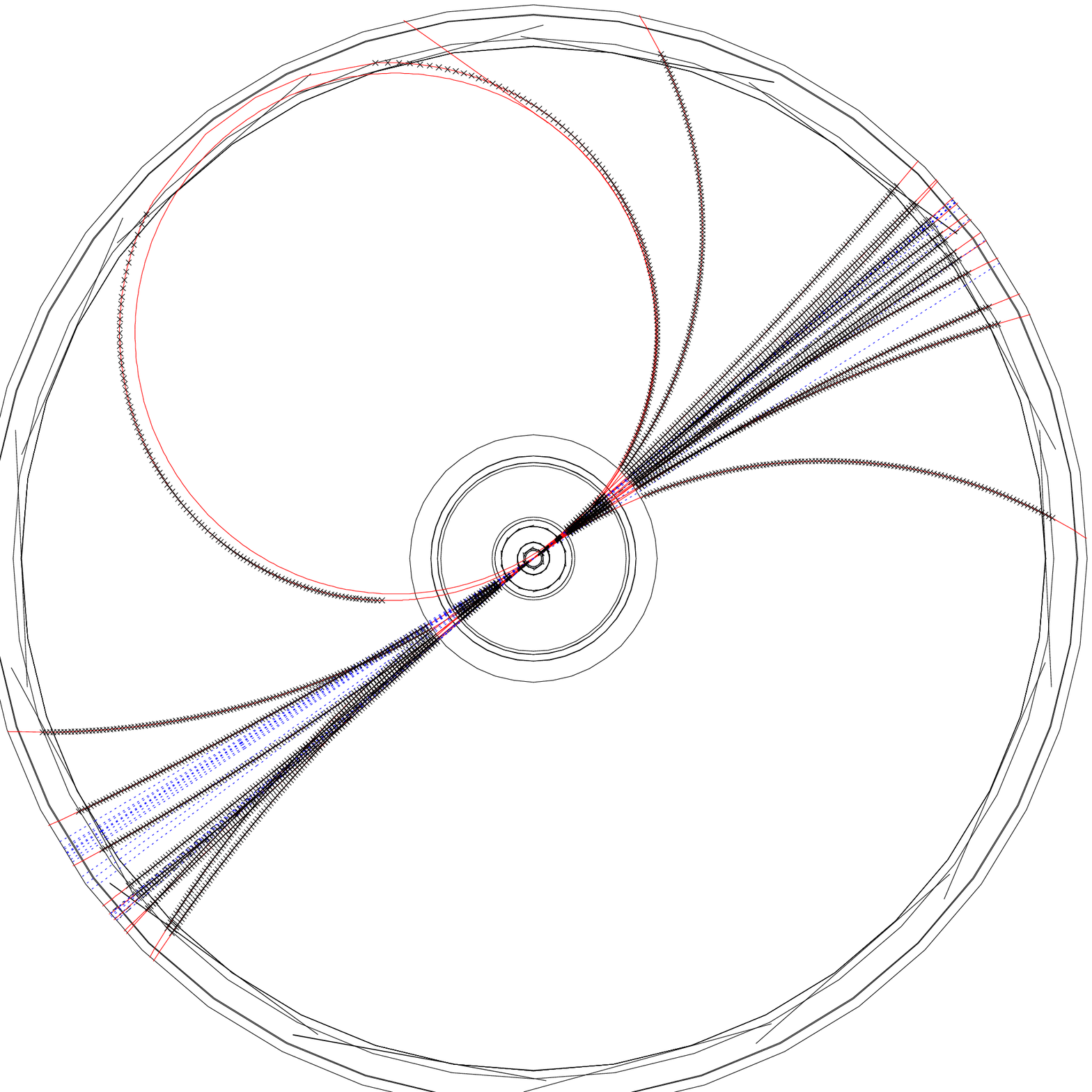,width=5.5cm} &
\epsfig{file=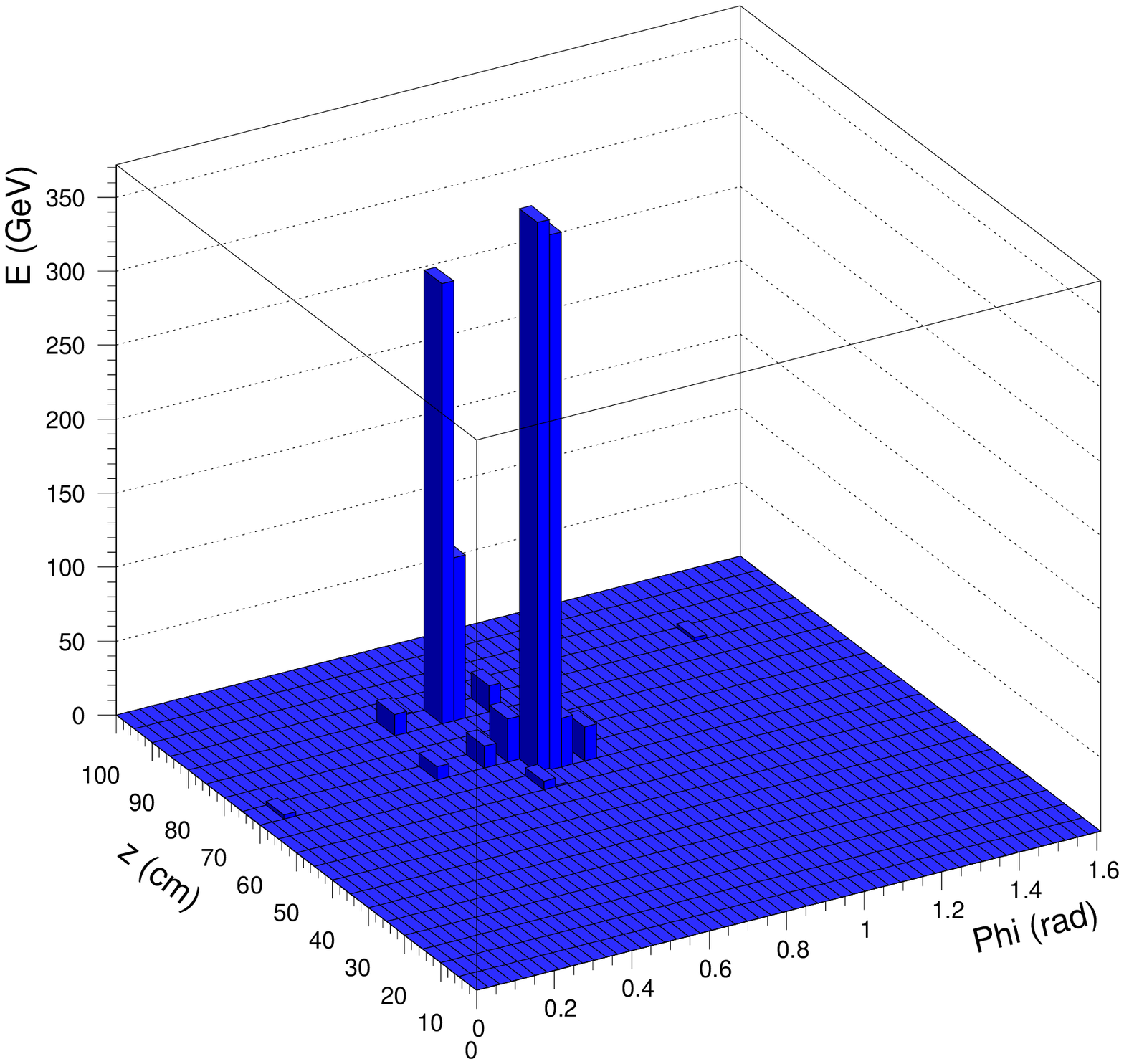,width=6.5cm} \\
\end{tabular}
\vspace*{-0.75cm}
\end{center}
\caption{\sl A $e^+e^- \rightarrow W^+W^-$ event at $\sqrt{s}$ = 3~TeV (left)
and the energy flowing at the entrance of the e.m. calorimeter located at a 
radius of 170~cm assuming a solenoidal field of 6~T and 5 $\times$ 5 cm$^2$
cell size (right).}
\label{fig:wwjj}
\end{figure}

The anticipated large jet multiplicity and boost make it necessary to 
reconsider the definition of jet clustering, intra-jet and inter-jet particles.
In particular for processes like $e^+e^- \rightarrow W^+W^-$, with a 
cross-section of 0.5~pb at $\sqrt{s}$ = 3~TeV, the jet collimation may make
the resolution of the individual charged and neutral particles problematic.
Figure~\ref{fig:occup} shows the distance from the closest track for a 
charged particle with $p >$ 3~GeV/$c$ at different radial distances from the 
interaction point, while Figure~\ref{fig:wwjj} shows the energy flow at the 
entrance of a calorimeter located at a radius of 170~cm for a typical 
$W^+W^-$ hadronic decay. 

These considerations seem to favour an inclusive approach where hadronic jets 
are reconstructed as elementary objects in the calorimeters.
On the other hand, jet pairing ambiguity in channels with large jet
multiplicity and heavy quark identification make jet flavour and charge
identification highly important in order to suppress the combinatorial and 
multi-fermion backgrounds. This in turn requires efficient and accurate 
tracking capabilities in collimated hadronic jets.

\subsection{Missing Energy and Forward Processes}

The detector ability to identify events with large missing energy is an 
important requisite since this is the main signature for production and 
decay of supersymmetric particles, such as $\tilde \ell^+ \tilde \ell^-$ and
$\tilde \chi^+ \tilde \chi^-$, and a possible complement in the study
of the Higgs sector through invisible decays. 
\begin{figure}[h!]
\begin{center}
\vspace*{-0.75cm}
\epsfig{file=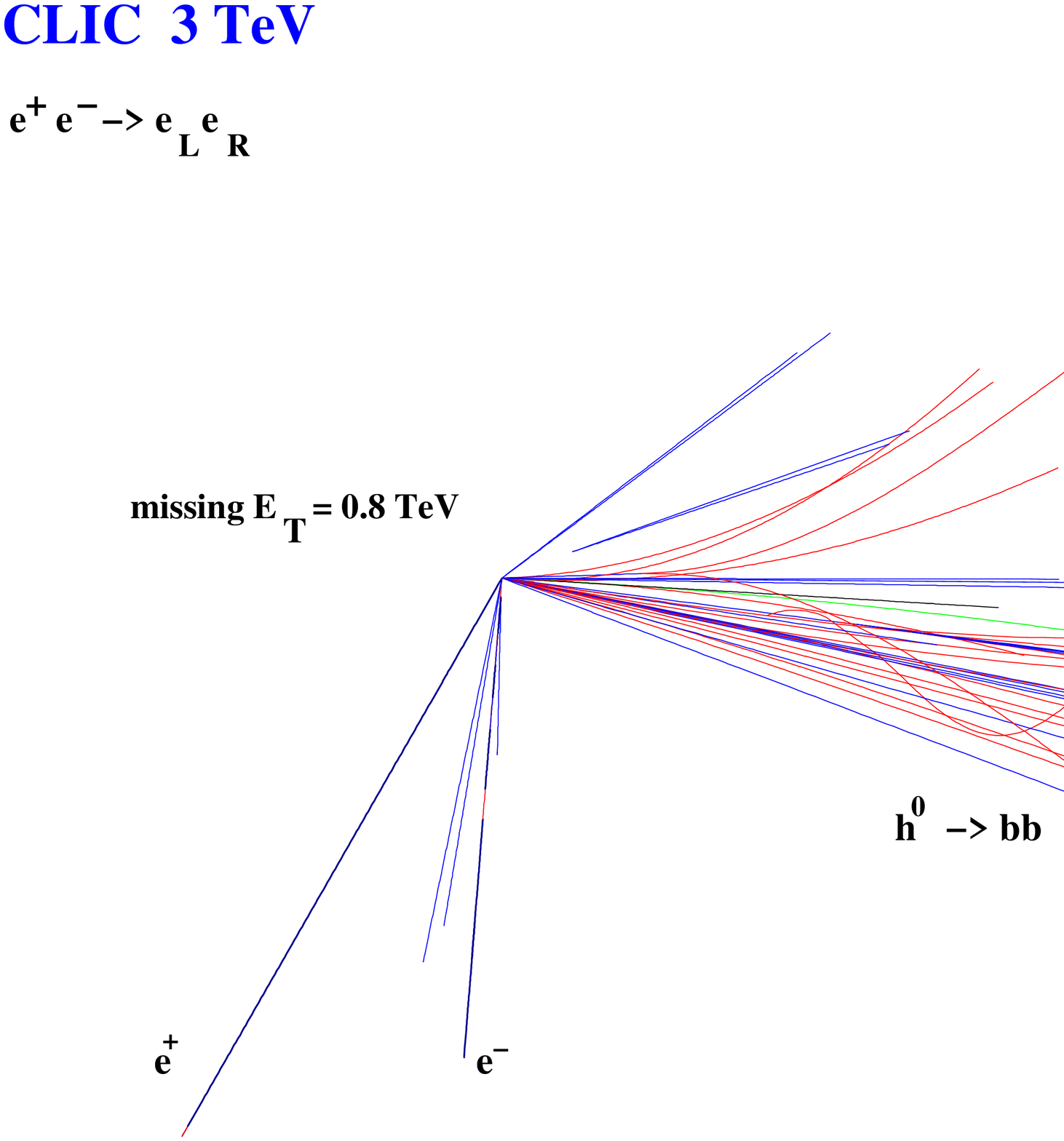,bbllx=10,bblly=90,bburx=650,bbury=760,width=7.0cm}
\vspace*{-0.75cm}
\end{center}
\caption{\sl Event display of a $e^+e^- \rightarrow \tilde e_L \tilde e_R$ 
reaction followed by the cascade decays
$\tilde e_R \rightarrow e \chi^0_1$, $\tilde e_L \rightarrow e \chi^0_2$,
$\chi^0_2 \rightarrow h \chi^0_1$, $h \rightarrow b \bar b$ for selectron 
masses of 1050 GeV and  $m_h$ = 115 GeV/$c^2$.}
\label{fig:eler}
\end{figure}
In particular the search for
$e^+e^- \rightarrow \tilde \ell^+ \tilde \ell^-$ at $\sqrt{s} >$ 1~TeV
will fully cover the region of the $(m_{1/2},m_0)$ MSSM parameter plane 
preferred by the LSP cosmological relic density~\cite{dm}.
 
Processes whose production cross section is peaked in the forward region,
such as Higgs production in $WW$ and $ZZ$ fusion, are also important and their
observability must be guaranteed. 
Finally, the study of $W^+W^- \rightarrow W^+W^-$ and $W^+W^- \rightarrow 
Z^0 Z^0$ cross sections probes the possible dynamics of strong electro-weak
symmetry breaking at $\sqrt{s} >$ 1~TeV~\cite{barger}. 
In order to ensure an efficient rejection of the $e^+e^-W^+W^-$ and $e^{\pm} 
\nu W^{\pm} Z^0$ backgrounds, electron detection capabilities must be 
ensured down to small polar angles. 

\begin{figure}[h!]
\begin{center}
\vspace*{-0.75cm}
\begin{tabular}{c c}
\epsfig{file=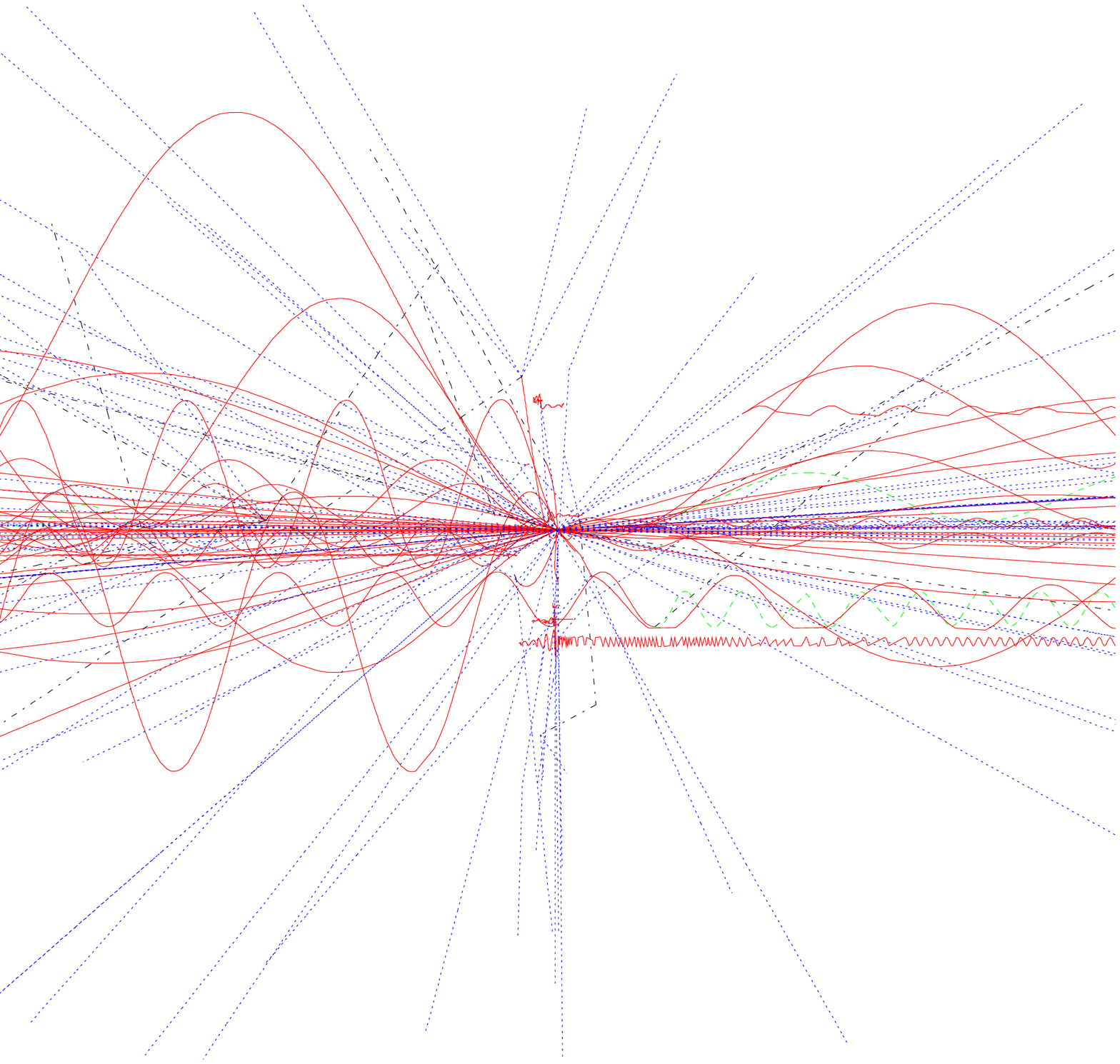,width=6.0cm} &
\epsfig{file=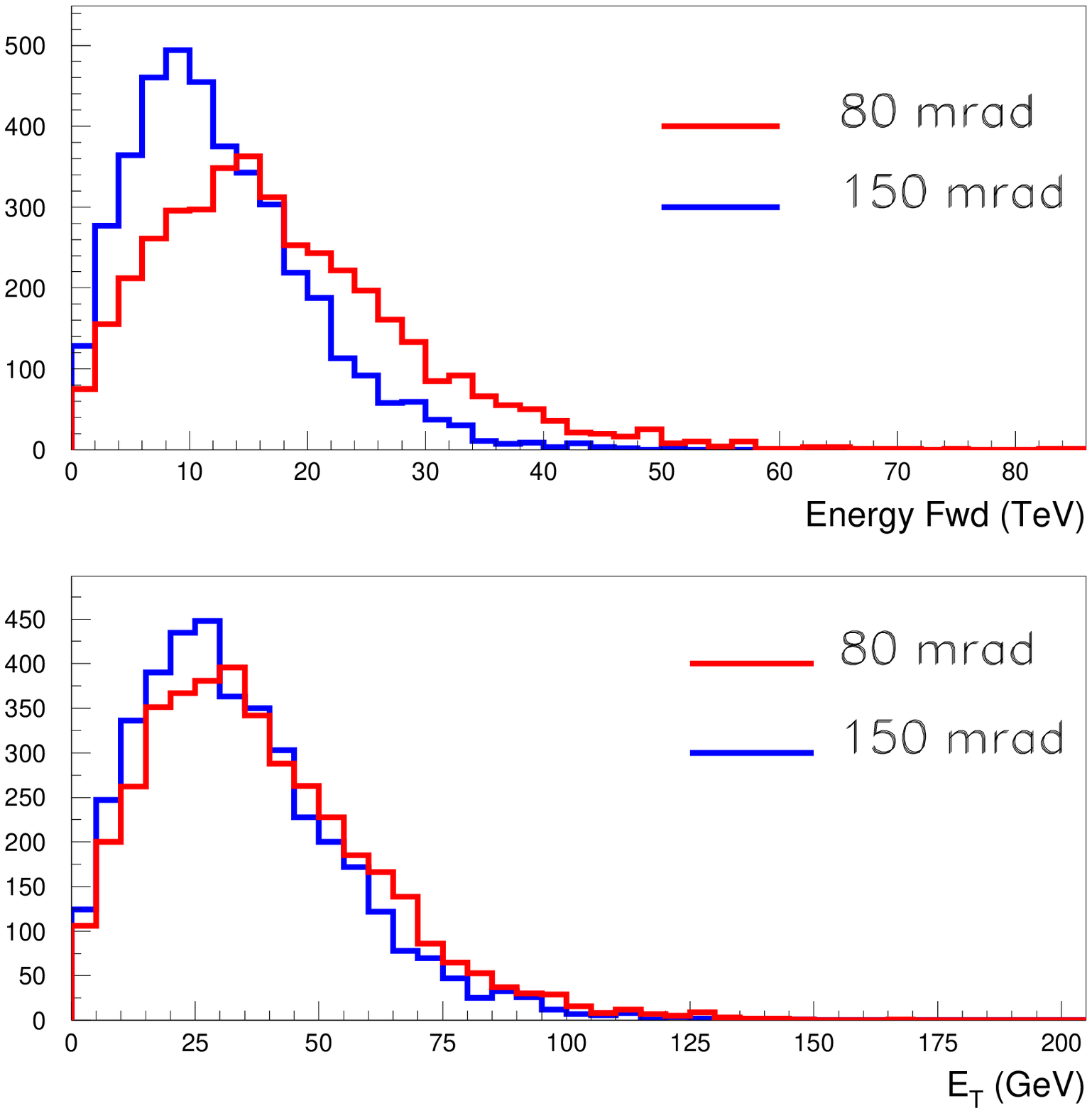,width=7.0cm,height=6.5cm} \\
\end{tabular}
\vspace*{-0.75cm}
\end{center}
\caption{\sl Event display (left) and $E$ and $E_T$ distributions 
(right) for the $\gamma \gamma$ background events generated in a single 
{\sc Clic} bunch crossing for different detector polar angle coverages.}
\label{fig:gg}
\end{figure}

These requisites have to be verified against 
the significant $\gamma \gamma$ background that limits the event reconstruction
at small angles. The effect of the $\gamma \gamma$ background has been studied
by tracking the $\gamma \gamma \rightarrow$ hadrons products using the full 
{\sc Geant} simulation and recording the measured $E$ and $E_T$ assuming a 
detector polar angle coverage down to 150 or 80~mrad. The results are shown
in Figure~\ref{fig:gg}. It is observed that, while the energy deposition
is very significant, the observed transverse energy is limited at about 
100~GeV. 

\begin{figure}[h!]
\begin{center}
\vspace*{-0.75cm}
\begin{tabular}{c c}
\epsfig{file=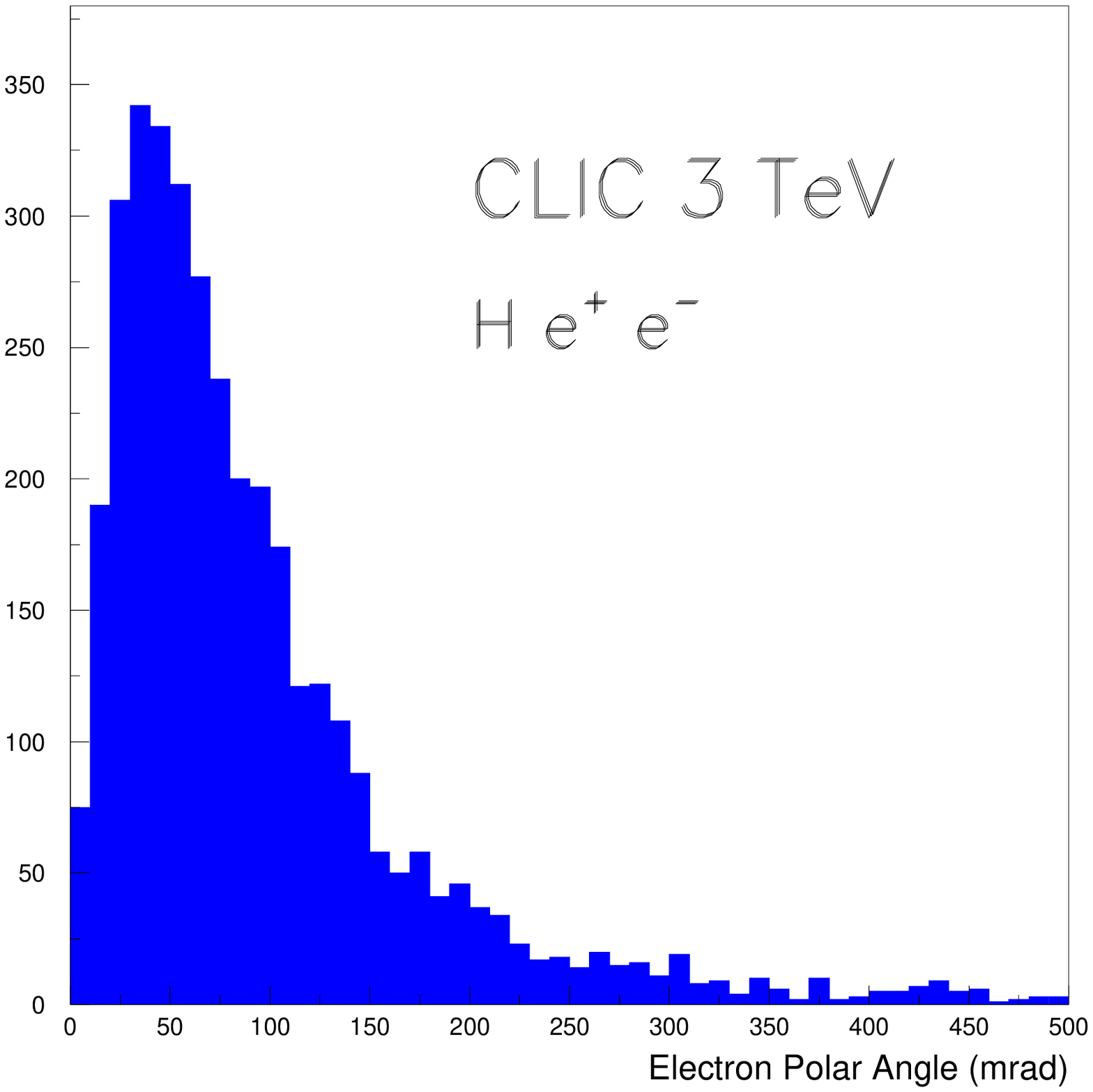, width=6.5cm, height=6.0cm} &
\epsfig{file=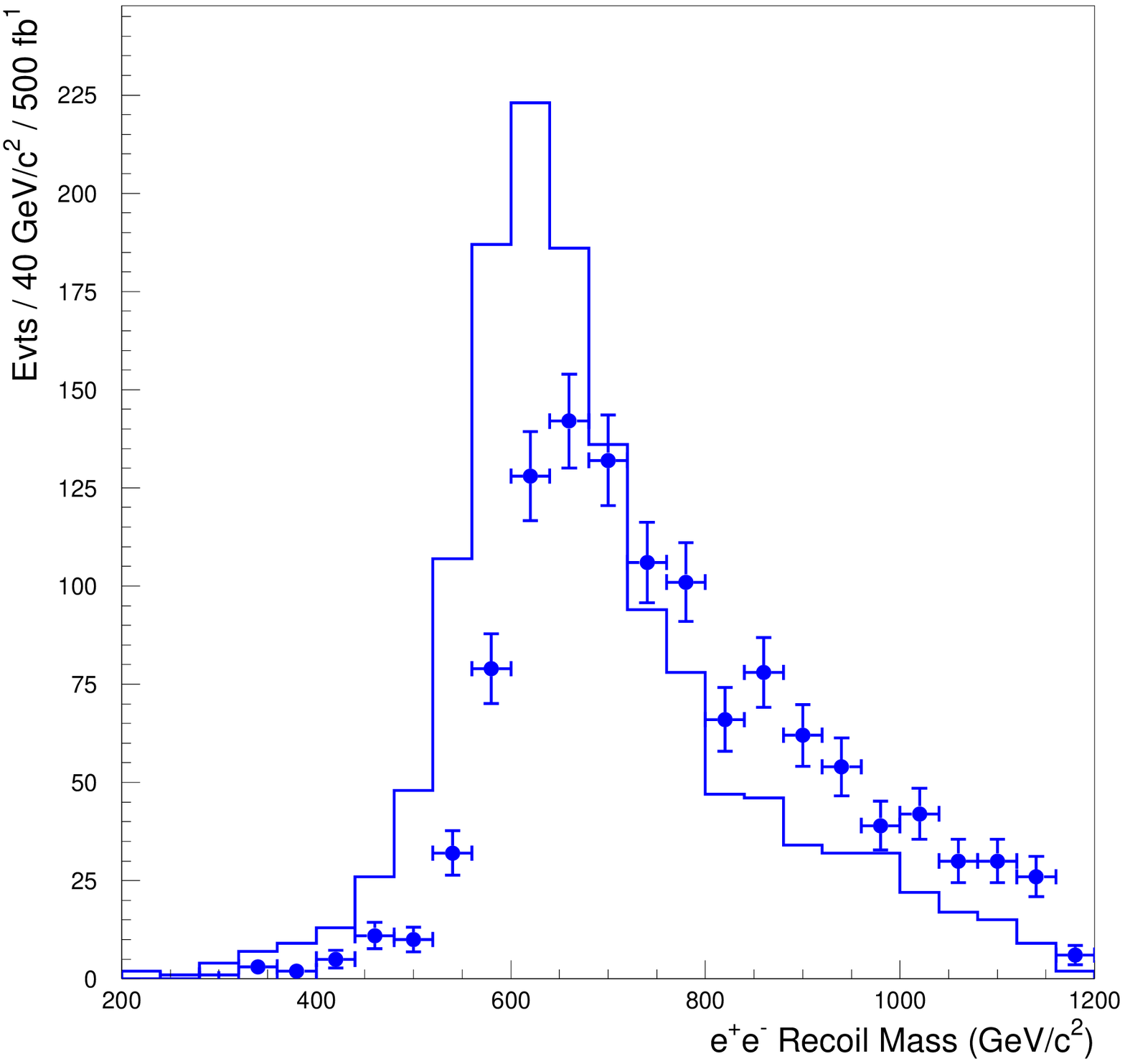, width=6.5cm, height=6.0cm} \\
\end{tabular}
\vspace*{-0.75cm}
\end{center}
\caption{\sl Polar angle distribution for electrons (left) and $ee$ recoil
mass spectrum in the $ZZ$ fusion reaction $e^+e^- \rightarrow H^0 e^+e^-$ for 
$\sqrt{s}$ = 3~TeV and $M_H$ = 600~GeV/$c^2$. The histogram (points with
error bars) show the distribution before (after) accounting for the luminosity
spectrum and $\gamma \gamma$ background.}
\label{fig:eeh1}
\end{figure}

Higgs production via $ZZ$ fusion in the process 
$e^+e^- \rightarrow H^0 e^+e^-$ has been investigated for $M_H$ = 
600~GeV/$c^2$. At high $\sqrt{s}$ and large values of the Higgs mass the 
$ZZ$ fusion process is favourable to study an exotic or invisibly decaying 
Higgs. If the $e^+ e^-$ in the final state can be tagged, this process 
allows to identify Higgs decays by the recoil mass independently of the 
Higgs decay modes and benefits of a production cross section of 30~fb to 
15~fb for Higgs masses in the range 400~GeV/$c^2$ to 900~GeV/$c^2$ compared
to $\simeq$~1.5~fb for the Higgs-strahlung $HZ$ process. 
Figure~\ref{fig:eeh1} shows the production polar angle for the electron and 
positron peaking at $\simeq$~50~mrad. Assuming tagging and tracking 
capabilities extending to 120~mrad, for electron with energy exceeding 
700~GeV, the $e^+e^-$ recoil mass can be used to isolate the Higgs signal 
also when accounting for the mass peak broadening due to the luminosity 
spectrum. 

\section{Conclusions}

A first survey of signatures for physics processes of potential interests at
the {\sc Clic} $e^+e^-$ linear collider at $\sqrt{s} \simeq$ 3~TeV has been
performed. Four physics signatures have been identified: i) Resonance Scan,
ii) Electro-weak Fits, iii) Multi-Jet Final States and iv) Missing Energy and 
Forward Processes and examples of reference reactions have been identified to
test the effect of the accelerator parameters and of the detector response on 
the {\sc Clic} physics potential.

\section*{Acknowledgements}

I am grateful to S.~De~Curtis, D.~Dominici, J.~Ellis, A.~De~Roeck, G.~Guignard,
D.~Schulte, R.~Settles, D.~Treille and G.~Wilson for ideas and suggestions 
on several topics discussed in this note. 

\newpage

\begin{table}[ht!]
\begin{center}

\caption{\sl Physics Signatures and CLIC Physics Program}

\vspace{0.5cm}

\begin{tabular}{|l||c|c|c|c|c|}

\hline
Physics        & Higgs  & SUSY & SSB & New Gauge   & Extra \\
Signatures     & Sector &      &     & Bosons      & Dimensions \\
\hline \hline

Resonance Scan &       & SUSY &  D-BESS  & $Z'$      & KK 
\\ 
               &       & Thresholds &    &         &    \\ \hline
EW Fits        &       &      &     & $A_{LR}$, &                     \\
               &       &      &     & $A_{FB}$  & $A_{FB}^{b \bar b}$ 
\\ \hline
Multi-Jets     & $H^+H^-$ &   & Techni-$\rho$  &           &  \\
               & $t \bar t H$ & &   &           &  \\        
               & $H H \nu \bar \nu$ &  &    &           &  \\        
               & $H H Z$  & &       &           &   \\
               & $H H H \nu \bar \nu$ &  &    &           &  \\        
               & $H H H Z$  & &       &           &

\\ \hline
$E_{miss}$, Fwd & $He^+e^-$ & $\tilde \ell^+ \tilde \ell^-$ & $WW$ &  
               &  \\
               &       & $\tilde \chi^+ \tilde \chi^-$ & scattering &   &
\\ \hline
\end{tabular}
\end{center}
\label{table:physprog}
\end{table}

\begin{table}[hb!]

\begin{center}

\caption{\sl Physics Signatures and CLIC Parameters: Benchmarks and Challenges}

\vspace{0.5cm}

\begin{tabular}{|l||c|c|c|c|c|}

\hline
Physics        & Beam-  & Beam & $e^+$ & Pairs & $\gamma \gamma$ Bkg. \\
Signatures     & Strahlung & $E$ Spread & Polarisation   &     &     \\
\hline \hline

Resonance & Stat.      &  Shape & Couplings  & $\Gamma_{bb}, _{cc}, _{tt}$ &
$\Gamma_{bb}, _{cc}, _{tt}$ \\
Scan               & Shape Syst. & Syst.      &         &   &        
\\ \hline
EW Fits        & Unfold & & Polarisation & $b \bar b$, $c \bar c$ & 
$\cos \theta_{min}$\\
               & Boost & & Measurement & Tags & Bkg Flavour      
\\ \hline
Multi-Jets     & 5-C Fit &  & & Tags for     & Fake Jets  \\               
                  &   & &  & Jet pairing  &  \\ \hline
$E_{miss}$ & $\theta_{miss}$ & & & Fwd  & $E_{hem}$  \\
Fwd        &  & & & Tracking & $E_T$
\\ \hline
\end{tabular}
\end{center}
\label{table:physpar}
\end{table}

\end{document}